\documentclass[conference]{IEEEtran}

\IEEEoverridecommandlockouts
\usepackage[bookmarks=false]{hyperref}
\usepackage{microtype}
\usepackage{subfigure}
\usepackage{booktabs} 
\usepackage{color, colortbl}
\usepackage{tablefootnote}
\usepackage[export]{adjustbox}
\usepackage{amsmath,amssymb,amsfonts}
\usepackage{algorithmic}
\usepackage{graphicx}
\usepackage{textcomp}
\usepackage{xcolor}
\usepackage{cite}

\def\BibTeX{{\rm B\kern-.05em{\sc i\kern-.025em b}\kern-.08em
    T\kern-.1667em\lower.7ex\hbox{E}\kern-.125emX}}
\definecolor{Gray}{gray}{0.9}
\graphicspath{{resources/}} 
\DeclareGraphicsExtensions{.pdf,.jpeg,.png} 
\newcommand{\reals}{\rm I\!R}

\begin{document}

\title{Benchmarking Network Embedding Models for Link Prediction: Are We Making Progress?\\
	\thanks{The research leading to these results has received funding from the European Research Council under the European Union's Seventh Framework Programme (FP7/2007-2013) / ERC Grant Agreement no. 615517, from the Flemish Government under the ``Onderzoeksprogramma Artificiële Intelligentie (AI) Vlaanderen'' programme, and from the FWO (project no. G091017N, G0F9816N, 3G042220).}
}

\author{\IEEEauthorblockN{1\textsuperscript{st} Alexandru Cristian Mara}
	\IEEEauthorblockA{\textit{Dept. of Electronics and I. S.} \\
		\textit{Ghent University}\\
		Ghent, Belgium \\
		alexandru.mara(at)ugent.be}
	\and
	\IEEEauthorblockN{2\textsuperscript{nd} Jefrey Lijffijt}
	\IEEEauthorblockA{\textit{Dept. of Electronics and I. S.} \\
		\textit{Ghent University}\\
		Ghent, Belgium \\
		jefrey.lijffijt(at)ugent.be}
	\and
	\IEEEauthorblockN{3\textsuperscript{rd} Tijl de Bie}
	\IEEEauthorblockA{\textit{Dept. of Electronics and I. S.} \\
		\textit{Ghent University}\\
		Ghent, Belgium \\
		tijl.debie(at)ugent.be}
}

\maketitle

\begin{abstract}
Network embedding methods map a network's nodes to vectors in an embedding space, in such a way that these representations are useful for estimating some notion of similarity or proximity between pairs of nodes in the network. The quality of these node representations is then showcased through results of downstream prediction tasks. Commonly used benchmark tasks such as link prediction, however, present complex evaluation pipelines and an abundance of design choices. This, together with a lack of standardized evaluation setups can obscure the real progress in the field. In this paper, we aim to shed light on the state-of-the-art of network embedding methods for link prediction and show, using a consistent evaluation pipeline, that only thin progress has been made over the last years. The newly conducted benchmark that we present here, including 17 embedding methods, also shows that many approaches are outperformed even by simple heuristics. Finally, we argue that standardized evaluation tools can repair this situation and boost future progress in this field.
\end{abstract}


\section{Introduction}\label{sec_intro}

Network representation learning or network embedding (NE) methods have attracted much interest in recent years \cite{belkin2002laplacian,perozzi2014deepwalk,tang2015line,cao2015grarep,ou2016hope,Gao2018bine}. These methods aim to bridge the gap between network structured data and traditional machine learning by constructing low-dimensional embeddings of network nodes as vectors, for example in the metric space. Using these embeddings, traditional machine learning approaches can be used to perform inference or obtain predictions from network data. A prominent application of NEs is Link Prediction (LP) \cite{grover2016node2vec,PRUNE2017,kang2018cne}, which amounts to estimating the probability of the existence of edges between nodes not connected in the input network.

In other tasks such as multi-label classification, a general consensus on the evaluation setup and criteria exists \cite{william2017rl, zhang2018nrl, Goyal18}, but for LP this is not the case. The evaluation of LP performance requires a pipeline with several preprocessing steps and design choices that can confound the results and are prone to errors. The main challenges for LP evaluation are:

\emph{1) Data preprocessing:} To assess the generalization performance of an NE method for LP, sets of train and test edges (i.e connected node-pairs) are required. These sets can be obtained using different approaches that may impact the evaluation in various ways. For instance, a typical assumption in LP is that the train edges span a (train) sub-network of a more complete (test) network. Thus, a principled approach is to use snapshots of the network at two different points in time. The first snapshot is used for training and the difference between the two snapshots for testing \cite{lpFairEval2002, Yang2015elp, Garcia2015}. Unfortunately, networks with such a temporal component are scarce and therefore, authors ordinarily resort to \emph{sampling edges} from a network as test examples and using the remaining edges for training \cite{grover2016node2vec,PRUNE2017,Gao2018bine,kang2018cne}. 
The sampling process varies between scientific works in different aspects. The sizes of the train and test sets vary between 50-50 in \cite{grover2016node2vec,kang2018cne}, to 60-40 in \cite{Gao2018bine} and 80-20 in \cite{PRUNE2017}. The algorithms used to generate these splits also differ and while some aim to construct training networks with similar, scaled-down, properties such as, e.g., node degrees \cite{Gurukar2019}, others generate training graphs that preserve the general topology of the original network \cite{mara2019evalne}.
 
\emph{2) Prediction pipeline:} Different NE methods require different pipelines to obtain link predictions. While some embedding methods directly compute link probabilities \cite{Zhang2018arope, kang2018cne}, for others, these need to be learned on top of the node embeddings. Two common approaches are: (i) interpreting some notion of similarity between two node embeddings (e.g. dot product) as the link probability and (ii) casting the problem as a binary classification task. The latter, which has been shown to be more effective \cite{Gurukar2019}, requires as a pre-step the computation of node-pair embeddings. Thus, an operator must be applied on the node embeddings to obtain node-pair representations that are, in turn, fed into a binary classifier to perform LP. The choice of operator not only varies between scientific works \cite{grover2016node2vec, Tsit2018verse} but is, in some cases, completely overlooked in the experimental setup discussion \cite{wang2016sdne,PRUNE2017}. Moreover, many valid choices exist for the binary classifier, and classifier training requires---in addition to `positive' examples or edges---`negative' samples or non-edges. These non-edges can also be selected using different strategies and be of varying sizes \cite{2017negsanp}.
  
\emph{3) Hyperparameter tuning:} When comparing new methods with the state-of-the-art, for the baseline methods the default parameter settings are often used \cite{Zhang2018arope, Tsit2018verse}, yet care is taken in tuning the parameters of the introduced method. When the recommended default settings were informed by experiments on other graphs than those used in the study at hand, this can paint an unduly unfavourable picture of the baseline methods. 

\emph{4) Evaluation metrics:} Finally, no consensus exists on which evaluation criteria should be used for comparing different methods. While some papers advocate for the use of precision@$Np$ for a range of $Np$ values \cite{ou2016hope, wang2016sdne, Zhang2018arope}, others use AUC-ROC \cite{grover2016node2vec, kang2018cne} or precision and recall at fixed thresholds \cite{Wei2017nrcl}.

These challenges have led to an inconsistency in evaluation procedures throughout papers. Hence, the practical performance of NE methods for LP is poorly understood. 
Moreover, as several recent studies have found \cite{Hutson927,recsys19,blalock2020state}, the inconsistency of evaluation procedures is a central issue that has stalled progress in many sub-fields within AI. As such, in this paper we aim to clarify the impact of these design choices on the evaluation pipelines and ultimately on method performance. At the same time, we aim to shed light on the real state-of-the-art by means of a standard evaluation pipeline. 

\textbf{Contributions.} We provide an in-depth empirical analysis of the state-of-the-art on network representation learning for LP. In our experiments, we show that embedding-based LP methods can be matched in performance and sometimes even outperformed by simple heuristics. We also study the effect on performance of hyperparameter tuning, embedding dimensionality, train set size, edge sampling strategy, node-pair embedding operator, and classification methods. Inevitably there are also limitations on the scope of the evaluation, such as an exclusive focus on undirected and unweighted networks without attributes, an analysis of moderate-sized networks, and a representative but finite set of evaluated methods and benchmark networks. However, our evaluation employs an open source framework, \href{https://github.com/Dru-Mara/EvalNE}{\texttt{EvalNE}} \cite{mara2019evalne}, that ensures reproducibility of results and allows for direct extensions of this work in all the aforementioned aspects.

The paper is organized as follows. Section~\ref{sec_methods} describes the NE methods evaluated, Section~\ref{sec_datasets} presents the datasets used, Section~\ref{sec_setup} discusses the evaluation setup, Section~\ref{sec_results} summarizes the results, and Section~\ref{sec_conclusions} concludes this paper. 

\section{Methods}\label{sec_methods}

In this section we present the evaluated NE methods and baseline LP heuristics. When available, we compare implementations of NE methods from distinct sources. Specifically, in addition to original implementations we consider those in two actively maintained libraries, i.e. \href{https://github.com/thunlp/OpenNE}{\texttt{OpenNE}} and \texttt{GEM} \cite{Goyal18}. Table \ref{table:methods} summarizes the hyperparameters tuned and implementations used. In the table all LP heuristics are summarized in a single \textit{Heuristics} field. Detailed hyperparameter descriptions are provided in the  \href{https://bitbucket.org/ghentdatascience/nrl4lp-public/src/master/}{supplementary material}. For all methods where this is relevant, the node-pair embedding operator is tuned as an additional hyperparameter (see Section~\ref{sec_setup}). 

As for the notation, we represent an undirected network or graph as $\mathbf{G}=({\mathbf{V},\mathbf{E}})$ with vertex set $\mathbf{V}=\{1,\dots,N\}$ and edge set $\mathbf{E} \subseteq {\mathbf{V} \choose 2}$. Edges or connected node-pairs are represented as unordered pairs $\{i,j\} \in \mathbf{E}$. Non-edges or disconnected pairs are represented as $\{i,j\} \in \mathbf{D}$.
We refer to the sets of train node-pairs as $\mathbf{E}_{train}$ and $\mathbf{D}_{train}$ and to the test node-pair sets as $\mathbf{E}_{test}$ and $\mathbf{D}_{test}$. The adjacency matrix of a graph $\mathbf{G}$ is represented as $\mathbf{A}$. We use $\mathbf{X}= (\mathbf{x}_1, \mathbf{x}_2, \dots, \mathbf{x}_N)$ with $\mathbf{X} \in \reals^{N\times d}$ to denote a $d$-dimensional node embedding and $\Gamma(i)$ to refer to the neighbourhood of node $i$.

\subsection{Network Embedding Methods}

\begin{table}
	\centering
	\caption{Hyperparameters tuned and implementations evaluated for each method. Except for AROPE, CNE and the LP heuristics, we also tune the edge embedding operator as a hyperparameter.}
	\label{table:methods}
	\vskip 0.1in
	\resizebox{\columnwidth}{!}{
		\begin{tabular}{lll}
			\toprule
			Methods	&Implementations &Parameters	 \\ \hline
			Heuristics	& \texttt{EvalNE} & -		\\
			DeepWalk		& Orig., \texttt{OpenNE} & $num\_ walks=walk\_ len=[5,10,20,40,80]$, \\ 
			& & $window\_ size=[5, 10, 20]$	\\
			Node2vec		& Orig., \texttt{OpenNE} & $num\_ walks=walk\_ len=[5,10,20,40,80]$, \\
			& & $window\_ size=[5, 10, 20]$, $p=q=[0.5, 1, 2]$ 	 \\
			Struc2vec		& Orig. & $num\_ walks=walk\_ len=[5,10,20,40,80]$, \\
			& & $window\_ size=[5, 10, 20]$	 \\
			Metapath2vec	& Orig. & $\alpha=[0.01, 0.025]$, $negative=[5, 10]$ \\
			WYS		& \href{https://github.com/benedekrozemberczki/AttentionWalk}{Other} & $lr=[0.01, 0.05]$, $num\_ walks=[20,40,80]$, \\
			& & $window\_ size=[5, 10, 20]$ \\ 
			GF 		& \texttt{OpenNE} & -	 \\
			GraRep 	& \texttt{OpenNE} & $kstep=[2, 4, 8]$	 \\
			HOPE 	& \texttt{OpenNE}, \texttt{GEM} & $\beta=[0.1, 0.01, 0.001, 0.0001]$ \\
			LE 		& \texttt{OpenNE}, \texttt{GEM} & -	 \\
			LLE 	& \texttt{GEM} & - \\
			M-NMF 	& Orig. & $clusters=[10, 20, 50]$	\\
			AROPE 	& Orig. & $weights=[[1,0,0,0], [0,1,0,0], [0,0,1,0]$, \\ 
			& & $[0,0,0,1], [1,0.1,0.01,0.001], [1,0.5,0.05,0.005]]$ \\
			SDNE	& \texttt{OpenNE}, \texttt{GEM} & $\beta=[2, 5, 10]$, \\ 
			& & $encoder\_list=[[128], [512,128], [1024,512,128]]$ \\
			PRUNE	& Orig. & $\lambda=[0.01, 0.05]$	\\
			VERSE	& Orig. & $nsamples=[3, 5, 10]$	 \\
			LINE	& Orig. \texttt{OpenNE} & $\rho=[0.01, 0.025]$, $negative\_ratio=[5, 10]$	\\
			CNE		& Orig. & $lr=[0.01, 0.05]$ \\
			\bottomrule
		\end{tabular}
	}
\end{table}

NE methods can be broadly categorized into four classes according to the strategy used for learning node similarities. In this taxonomy we can distinguish methods based on random walks, matrix factorization, neural networks, and probabilistic approaches. Next, we describe the methods from each family included in our evaluation. 

\subsubsection{Methods based on random walks} \label{ss_rw}
These methods determine node similarities using random walks on the input graph. The Skip-Gram model \cite{mikolov2013skipgram}, is then commonly used to generate node embeddings from the random walks.

\emph{DeepWalk} \cite{perozzi2014deepwalk} is the first method to use deep learning inspired techniques for NE. It uses random walks with fixed transition probabilities to measure node similarities and embeddings are derived using the Skip-Gram model approximated via hierarchical softmax. 

\emph{Node2vec} \cite{grover2016node2vec} is a generalization of DeepWalk which uses truncated random walks for node neighbourhood exploration and the Skip-Gram model, approximated via negative sampling, for embedding generation. The random walk properties are controlled by a return parameter $p$ and an in-out parameter $q$. 

\emph{Struc2vec} \cite{ribeiro2017struc2vec} extracts structural information from graphs through node pair similarities for a range of neighbourhood sizes. This information is then summarized as a multi-layer weighted graph $\mathbf{\hat{G}}$. Subsequently, random walks on $\mathbf{\hat{G}}$ are used to generate the embeddings. 

\emph{Metapath2vec} \cite{Dong2017metapath2vec} is a method capable of learning embeddings from heterogeneous networks. The authors extend the concept of random walks to account for nodes of different types and further use a heterogeneous Skip-Gram model to learn the embeddings. 

\emph{Watch Your Step (WYS)} \cite{abu2017graphatt} is an attention model learned on the power series of the transition matrix of $\mathbf{G}$. Node context importance is learned with minimal manually-tunable hyperparameters. 

\subsubsection{Methods based on matrix factorization} \label{ss_mf}
These approaches use representations of node similarities such as high-order proximities expressed as polynomials of $\mathbf{A}$, the incidence matrix, Katz similarity or the graph Laplacian. Node embeddings are then obtained by factorizing the selected matrix. 

\emph{Graph Factorization (GF)} \cite{ahmed2013distributed} uses regularized Gaussian matrix factorization to recover a matrix $\mathbf{Z}$ such that $\mathbf{Z}\mathbf{Z}^T$ is close to $\mathbf{A}$ in terms of observed non-zeros.

\emph{GraRep} \cite{cao2015grarep} is based on factorization of different polynomials of the adjacency matrix $\mathbf{A}$ which identify relations between nodes in $\mathbf{G}$ at different resolutions. 

\emph{HOPE} \cite{ou2016hope} also based on matrix factorization and preserving high-order proximities in graphs, additionally accounts for the asymmetric transitivity property of directed networks. 

\emph{Laplacian Eigenmaps (LE)} \cite{belkin2002laplacian} first constructs a weighted representation $\mathbf{\hat{A}}$ of $\mathbf{A}$ by leveraging first order proximities on $\mathbf{G}$. The Laplacian matrix $\mathbf{L}$ is computed using $\mathbf{\hat{A}}$ and embeddings are obtained from the $d$ eigenvectors corresponding to the lowest eigenvalues of $\mathbf{L}$. 

\emph{Locally Linear Embeddings (LLE)} \cite{Roweis2000lle} operates under the assumption that nodes and their neighbours lie on locally linear patches of a high-dimensional manifold. The embedding of a node can, thus, be derived from the linear coefficients that better reconstruct the node from the embeddings of its neighbours. 

\emph{M-NMF} \cite{Wang2017mnmf} incorporates community structure information in the embedding learning process via modularized non-negative matrix factorization. 

\emph{AROPE} \cite{Zhang2018arope} similarly to GraRep, proposes embeddings as found by the truncated singular value decomposition of polynomials of $\mathbf{A}$. The authors describe a fast eigen-decomposition method for these polynomials based on shifting or reweighing the decomposition of $\mathbf{A}$. 

\subsubsection{Methods based on neural Networks} \label{ss_nn}
Due to their ability to capture highly non-linear relations, deep models and particularly auto-encoders have also been used for network representation learning. 

\emph{Structural Deep Network Embedding (SDNE)} \cite{wang2016sdne} uses a deep neural network for learning embeddings that capture first and second order proximities on the graph. 

\emph{PRUNE} \cite{PRUNE2017} relies on a deep siamese neural network for learning node embeddings and can incorporate node ranking as additional information. 

\emph{VERSE} \cite{Tsit2018verse} learns embeddings by training a single layer neural network that minimizes the Kullback-Leibler (KL) divergence between node similarities in $\mathbf{G}$ and vector similarities in $\mathbf{X}$. Noise Contrastive Estimation is used for scalability. 

\subsubsection{Probabilistic Methods} \label{ss_prob}
These methods use probabilistic approaches to model node similarities and learn embeddings. 

\emph{LINE} \cite{tang2015line} uses joint and conditional probability distributions to model the first and second order adjacencies between linked nodes in $\mathbf{G}$. The Skip-Gram model is used to obtain node embeddings. 

\emph{CNE} \cite{kang2018cne} uses a Bayesian approach to generate embeddings which model the observed network while taking prior information into account. The prior can incorporate structural graph properties such as node degrees or block densities for clustered or multi-partite networks.

\subsection{Baseline Heuristics}\label{sec_heuristics}

In addition to the NE methods, we evaluate a set of LP heuristics as baselines. These heuristics use the neighbourhoods $\Gamma(i)$ and $\Gamma(j)$ of nodes pairs $\{i,j\}$ to derive similarity scores which can be interpreted as link probabilities. In our experiments we consider Common Neighbours defined as: $CN_{\{i,j\}} = |\Gamma(i) \cap \Gamma(j)|$; Jaccard Coefficient, $JC_{\{i,j\}} = |\Gamma(i) \cap \Gamma(j)|/|\Gamma(i) \cup \Gamma(j)|$; Adamic-Adar Index, $AA_{\{i,j\}} = \sum_{k \in \Gamma(i) \cap \Gamma(j)} 1/\log |\Gamma(k)|$; Resource Allocation Index, $RAI_{\{i,j\}} = \sum_{k \in \Gamma(i) \cap \Gamma(j)} 1/|\Gamma(k)|$ and Preferential Attachment, $PA_{\{i,j\}} = |\Gamma(i)| \cdot |\Gamma(j)|$.

Additionally, we generate a `heuristics embedding' by concatenating CN, JC, AA, RAI, and PA as a 5-dimensional node-pair embedding. Logistic regression is then used to obtain link predictions. We will refer to this method as \textit{NE\_heuristics}. 

\section{Datasets}\label{sec_datasets}

We conduct our experimental evaluation on 7 undirected real-world networks from different domains. These networks are medium-sized to ensure successful execution of all methods and constrain the computational resources needed.
Next, we present a short description of each network and in Table \ref{table:datasets}, we summarize their main statistics. 

\textit{StudentDB} \cite{goethals2010} represents a snapshot of Antwerp University's relational student database. Nodes in the network represent entities such as students, professors, tracks, etc. and edges constitute binary relations, i.e. student-in-track, student-in-program, student-take-course, professor-teach-course, course-in-room.
\textit{Facebook} \cite{leskovec2015snap} and \textit{BlogCatalog} \cite{zafarani2009social} are online social networks where nodes represent different users and edges indicate friendships. 
\textit{GR-QC} \cite{leskovec2015snap} and \textit{AstroPh} \cite{leskovec2015snap} describe collaboration networks in the fields of General Relativity and Astrophysics. Nodes represent papers and edges denote citations between them. 
\textit{PPI} \cite{breitkreutz2007biogrid} is a biological protein-protein interaction network and constitutes a subset of the Homo Sapiens PPI network. Finally,
\textit{Wikipedia} \cite{mahoney2011large} contains nodes representing words in Wikipedia pages and links denoting co-occurrences. 

\begin{table}
	\centering
	\caption{Summary of dataset statistics where $2|\mathbf{E}|/|\mathbf{V}|$ denotes the average node degrees of the undirected networks.}
	\label{table:datasets}
	\resizebox{\columnwidth}{!}{
		\begin{tabular}{lcccc}
			\toprule
			Dataset 		& Category & $|\mathbf{V}|$ & $|\mathbf{E}|$ & $2|\mathbf{E}|/|\mathbf{V}|$ \\ \hline
			\href{http://adrem.ua.ac.be/smurfig}{StudentDB}	& Relational & $395$   	& $3,423$ & $17.33$	\\
			\href{https://snap.stanford.edu/data/index.html}{Facebook}	& Social & $4,039$	& $88,234$  & $43.69$	\\
			\href{http://socialcomputing.asu.edu/pages/datasets}{BlogCatalog}	& Social & $10,312$ & $333,983$ & $64.77$	\\
			\href{https://snap.stanford.edu/data/index.html}{GR-QC}		& Collaboration & $4,158$ 	& $26,844$ & $6.45$	\\
			\href{https://snap.stanford.edu/data/index.html}{AstroPH}	& Collaboration & $18,772$ & $396,160$ & $22.00$	\\
			\href{http://snap.stanford.edu/node2vec}{PPI}   & Biological & $3,852$ 	& $37,841$ & $19.64$\\
			\href{http://snap.stanford.edu/node2vec}{Wikipedia}	& Language & $4,777$ 	& $92,295$ & $38.64$ \\
			\bottomrule
		\end{tabular}
	}
\end{table}

\section{Experimental Setup}\label{sec_setup}

In this section we give details on the LP task, present the experimental setup used and discuss the limitations to the scope of the evaluation and reproducibility of results.

\subsection{Link Prediction} \label{sec_lp}

As pointed out in Section \ref{sec_intro} the objective in LP is to identify missing links in an incomplete graph $\mathbf{G}$. Thus, the first step to evaluate LP is to preprocess $\mathbf{G}$ and obtain sets of train and test node-pairs. We use the standard approach of generating an incomplete training graph $\mathbf{G}_{train}=(\mathbf{V}, \mathbf{E}_{train})$ from a more complete graph $\mathbf{G}=(\mathbf{V}, \mathbf{E})$ where the connected node-pairs $\{i,j\} \in \mathbf{E} \setminus \mathbf{E}_{train}$ are used for testing. The proportion $f = |\mathbf{E}_{train}| / |\mathbf{E}|$ (or train fraction) is a user-defined parameter. For obtaining $\mathbf{E}_{train}$ and $\mathbf{E}_{test}$ we evaluate 3 approaches, namely \textit{random}  \cite{Gurukar2019}, \textit{spanning tree (ST)} \cite{mara2019evalne} and \textit{depth first tree (DFT)}. The first is a random sampling strategy followed by a main connected components computation, while ST and DFT constructively ensure that $\mathbf{G}_{train}$ contains a spanning tree of $\mathbf{G}$. Specifically, these strategies proceed as follows:
 
\subsubsection{Random} This approach proposed in \cite{Gurukar2019} starts by randomly selecting a set of edges from $\mathbf{E}$ as `preliminary' test edges, $\mathbf{\hat{E}}_{test}$. The main connected component spanned by the remaining edges $\{i,j\} \in \mathbf{E} \setminus \mathbf{\hat{E}}_{test}$ is then computed. These edges are considered as the training edges $\mathbf{E}_{train}$ spanning a training graph $\mathbf{G}_{train}=(\mathbf{V}_{train}, \mathbf{E}_{train})$. Finally, only the test edges $\{i, j\} \in \mathbf{\hat{E}}_{test}$ such that $i \in \mathbf{V}_{train}$ and $j \in \mathbf{V}_{train}$ form the set of final `refined' test edges $\mathbf{E}_{test}$. The sizes of the train and test edge sets are expected to vary between different executions of the algorithm. 

\subsubsection{ST} This approach proposed in \cite{mara2019evalne} first computes the main connected component of the input graph. Then, a spanning tree of this graph $\mathbf{G}$ is selected uniformly at random from the set of all possible ones using the algorithm proposed in \cite{wilson1996}. The edges in this spanning tree are then added to $\mathbf{E}_{train}$. Additional edges are selected uniformly at random from $\mathbf{G}$ and added to $\mathbf{E}_{train}$ until a set train edge fraction $f$ is reached. The remaining edges are used for testing. The sizes of the edge sets returned by this approach in different execution on the same graph are expected to be constant.

\subsubsection{DFT} This approach is a faster version of the ST method but with a fundamental difference i.e. the spanning tree of $\mathbf{G}$ is not selected uniformly at random, but computed using a deterministic depth first traversal algorithm. For small graphs this can result in highly similar sets of train edges obtained with different random seeds. Then, as for the ST approach, random edges are added to the spanning tree to form $\mathbf{E}_{train}$. The remaining edges are used for testing.

In addition to splitting connected pairs, we also generate sets of train and test non-edges, $\mathbf{D}_{train}$ and $\mathbf{D}_{test}$. The node-pairs in these sets are randomly selected using an open world assumption where any pair $\{i,j\} \notin \mathbf{E}_{train}$ is considered a valid train non-edge. Test non-edges are selected as pairs $\{i,j\} \notin (\mathbf{E} \cup \mathbf{D}_{train})$. In our experiments, we use the same number of connected and non-connected node-pairs.

For hyperparameter tuning the train sets need to be further split into train and validation. We fix these values to 90\% train and 10\% validation in all our experiments and the validation split is always performed using the same algorithm
as the initial train-test split. Grid search is adopted as the strategy for learning the best model hyperparameters.

For most methods, with the exception of CNE, AROPE and the LP heuristics, for which node-pair similarities are directly computed, the link predictions are learned through binary classification. First, node-pair embeddings for $\mathbf{E}_{train}$, $\mathbf{D}_{train}$, $\mathbf{E}_{test}$ and $\mathbf{D}_{test}$ need to be obtained from the node embeddings $\mathbf{X}$ learned by a method on $\mathbf{G}_{train}$. The embedding of a pair $\{i,j\}$ can be computed by applying different operators $\circ$ to the embeddings of the incident nodes $i$ and $j$ i.e. $\mathbf{x}_{\{i,j\}} = \mathbf{x}_i \circ \mathbf{x}_j$. In our evaluation, we use the operators introduced in \cite{grover2016node2vec}, namely \textit{Average} ($(\mathbf{x}_i + \mathbf{x}_j)/2$), \textit{Hadamard} ($\mathbf{x}_i \cdot \mathbf{x}_j$), \textit{Weighted $L_1$} ($|\mathbf{x}_i - \mathbf{x}_j|$) and \textit{Weighted $L_2$} ($|\mathbf{x}_i - \mathbf{x}_j|^2$). A binary classifier is then fitted with the node-pair embeddings and labels $\{0,1\}$ representing non-edges and edges, respectively.

\subsection{Evaluation Setup}

We use two experimental setups, called LP1 and LP2, with model hyperparameter tuning as only difference. While in LP1 we tune all hyperparameters presented in Table \ref{table:methods} as well as the node-pair operator, in LP2 we use default recommended parameters for all methods while still tuning the node-pair embedding strategy. In both cases, we provide results averaged over 3 experiment repeats and report test AUC-ROC values and execution times. We quantify method accuracy in terms of AUC-ROC as it is easily interpretable and the most commonly used LP evaluation metric. For visualization porpoises, we use $-\log(1-AUCROC)$, as it better reflects differences in performance for AUC-ROC values $\approxeq 1$. Unless otherwise specified we use embedding dimensionality $d=128$, train fraction $f=0.8$, ST as edge sampling strategy and logistic regression (with 5-fold CV to tune the regularization parameter) as binary classifier. Finally, all methods are run for their pre-defined number of iterations. 

Setup LP1 is used to analyse the performance of NE methods with respect to parameters $d \in \{8, 32, 128\}$ and $f \in \{0.2, 0.5, 0.8\}$. Setup LP2, on the other hand, is used to investigate the effect of the edge sampling strategies introduced in Section~\ref{sec_lp}, and the choice of binary classifier.

\subsection{Limits to the scope of the evaluation}

Constraints on overall computation time (the results in Section~\ref{sec_results} required 50.2 days to compute) inevitably imply boundaries to the scope of this evaluation: It is limited to popular mid-sized undirected, unweighted networks. It evaluates a representative but non-exhaustive set of state-of-the-art NE methods. It only evaluates methods purely exploiting the network structure (and not node/edge meta-data), such that some methods are not shown in their full potential (but note that the same holds for the heuristic baselines). Furthermore, the selected heuristics are very simple, other more powerful similarity metrics between network nodes have been proposed e.g. in \cite{Cukierski11,Hagberg08}. These are not included, as a thorough analysis of LP heuristic performance is not the main goal of our work. For a detailed overview and experimental evaluation of different heuristics we refer the reader to \cite{ghasemian2019stacking}. Including these more powerful predictors, however, could only strengthen the main conclusion drawn in this paper, that heuristic LP approaches, to date, remain competitive and in some cases outperform LP based on NE methods. 

\subsection{Reproducibility Notes}

In order to ensure the reproducibility of our results and foster further research in this area, we have based our experimental evaluation upon the \texttt{EvalNE} framework. This open source Python toolbox aims to simplify the evaluation of NE methods for LP, network reconstruction, node classification, and visualization. The toolbox automates tasks such as hyperparameter tuning, selection of train and test edges or non-edge sampling. It implements widely used node-pair embedding operators and can incorporate any classifier for prediction. Moreover, its design ensures that common errors, such as the computation of features on $\mathbf{G}$ rather than just on $\mathbf{G}_{train}$, or other forms of label leakage, are ruled out. Finally, \texttt{EvalNE} can assess the scalability of methods through wall clock time and performance through a wide range of accuracy measures e.g. AUC-ROC, PR, F-score, and threshold curves. Configuration files describing our complete evaluation setup and which can be used directly in \texttt{EvalNE} to replicate our results, are made available \href{https://bitbucket.org/ghentdatascience/nrl4lp-public/src/master/}{online}.

\section{Experimental Results}\label{sec_results}

\begin{table*}[!htbp]
	\centering
	\caption{AUC-ROC scores and standard deviations over 3 experiment repetitions for setup LP1 where hyperparameters are tuned and $d=128$ for all methods except CNE where $d=8$. Note that $0.000$ in the table means $<0.0005$. The best method within each type of approach is highlighted in bold and the overall best for each column on grey background.}
	\label{table:LP1}
	\vskip 0.1in
	\resizebox{\textwidth}{!}{
		\begin{tabular}{l|ccccccc|cc}
			\toprule
			Methods	&StudentDB	&Facebook	&BlogCat.	&GR-QC	&AstroPH	&PPI	&Wikipedia &Avg. AUC-ROC &Avg. Rank	\\ \hline 
			CN	&0.370$\pm$0.011	&0.992$\pm$0.001	&0.948$\pm$0.000	&0.959$\pm$0.001	&0.990$\pm$0.000	&0.863$\pm$0.003	&0.900$\pm$0.002	&0.860$\pm$0.210	&17.21 \\ 
			JC	&0.370$\pm$0.011	&0.990$\pm$0.001	&0.770$\pm$0.002	&0.959$\pm$0.001	&0.990$\pm$0.000	&0.839$\pm$0.001	&0.377$\pm$0.005	&0.756$\pm$0.260	&22.07 \\ 
			AA	&0.370$\pm$0.011	&0.993$\pm$0.001	&0.952$\pm$0.000	&0.959$\pm$0.001	&0.991$\pm$0.000	&0.867$\pm$0.003	&0.919$\pm$0.002	&0.864$\pm$0.211	&13.57 \\ 
			PA	&0.922$\pm$0.008	&0.842$\pm$0.003	&0.955$\pm$0.001	&0.839$\pm$0.006	&0.879$\pm$0.001	&0.905$\pm$0.002	&0.920$\pm$0.001	&0.894$\pm$0.041	&16.50 \\ 
			RAI	&0.370$\pm$0.011	&\textbf{0.994$\pm$0.001}	&\textbf{0.958$\pm$0.000}	&0.959$\pm$0.001	&0.991$\pm$0.000	&0.867$\pm$0.003	&\cellcolor{Gray}\textbf{0.931$\pm$0.002}	&0.867$\pm$0.212	&10.64 \\ 
			NE\_heuristics	&\textbf{0.966$\pm$0.004}	&0.993$\pm$0.000	&0.956$\pm$0.001	&\textbf{0.976$\pm$0.003}	&\textbf{0.993$\pm$0.000}	&\textbf{0.927$\pm$0.001}	&0.929$\pm$0.004	&\cellcolor{Gray}\textbf{0.963$\pm$0.026}	&\textbf{5.00} \\ \hline
			DeepWalk	&0.906$\pm$0.005	&0.990$\pm$0.000	&0.943$\pm$0.000	&\textbf{0.986$\pm$0.001}	&0.984$\pm$0.000	&0.905$\pm$0.001	&0.903$\pm$0.002	&0.945$\pm$0.040	&13.14 \\ 
			DeepWalk\_opne	&0.906$\pm$0.010	&0.991$\pm$0.000	&0.943$\pm$0.000	&0.985$\pm$0.002	&0.983$\pm$0.000	&\textbf{0.906$\pm$0.001}	&0.904$\pm$0.001	&\textbf{0.945$\pm$0.039}	&\textbf{13.00} \\ 
			Node2vec	&0.948$\pm$0.009	&\textbf{0.994$\pm$0.000}	&0.938$\pm$0.001	&0.985$\pm$0.003	&0.989$\pm$0.001	&0.840$\pm$0.006	&0.893$\pm$0.002	&0.941$\pm$0.054	&13.29 \\ 
			Node2vec\_opne	&0.897$\pm$0.004	&0.991$\pm$0.001	&0.929$\pm$0.001	&0.986$\pm$0.002	&\textbf{0.992$\pm$0.000}	&0.900$\pm$0.001	&0.901$\pm$0.001	&0.942$\pm$0.043	&13.57 \\ 
			Struc2vec	&0.933$\pm$0.010	&0.833$\pm$0.004	&\textbf{0.953$\pm$0.001}	&0.842$\pm$0.005	&0.874$\pm$0.001	&0.904$\pm$0.002	&\textbf{0.918$\pm$0.001}	&0.894$\pm$0.042	&18.00 \\ 
			Metapath2vec	&\textbf{0.981$\pm$0.005}	&0.942$\pm$0.003	&0.948$\pm$0.000	&0.804$\pm$0.006	&0.858$\pm$0.002	&0.880$\pm$0.003	&0.903$\pm$0.001	&0.902$\pm$0.058	&19.29 \\ 
			WYS	&0.819$\pm$0.016	&0.940$\pm$0.003	&0.915$\pm$0.002	&0.833$\pm$0.008	&0.855$\pm$0.004	&0.853$\pm$0.005	&0.864$\pm$0.010	&0.868$\pm$0.042	&25.57 \\  \hline
			GF\_opne	&0.868$\pm$0.007	&0.983$\pm$0.000	&0.898$\pm$0.001	&0.933$\pm$0.005	&0.947$\pm$0.001	&0.837$\pm$0.004	&0.834$\pm$0.003	&0.900$\pm$0.054	&23.57 \\ 
			GraRep\_opne	&0.969$\pm$0.003	&\textbf{0.993$\pm$0.000}	&\textbf{0.962$\pm$0.001}	&\textbf{0.984$\pm$0.002}	&\textbf{0.990$\pm$0.000}	&\textbf{0.921$\pm$0.001}	&\textbf{0.922$\pm$0.001}	&\cellcolor{Gray}\textbf{0.963$\pm$0.029}	&\textbf{5.29} \\ 
			HOPE\_gem	&\cellcolor{Gray}\textbf{0.989$\pm$0.001}	&0.990$\pm$0.000	&0.955$\pm$0.000	&0.952$\pm$0.002	&0.950$\pm$0.001	&0.909$\pm$0.002	&0.919$\pm$0.001	&0.952$\pm$0.029	&11.29 \\ 
			HOPE\_opne	&0.914$\pm$0.002	&0.989$\pm$0.000	&0.944$\pm$0.000	&0.920$\pm$0.005	&0.947$\pm$0.000	&0.872$\pm$0.005	&0.916$\pm$0.001	&0.929$\pm$0.034	&18.57 \\ 
			LE\_gem	&0.906$\pm$0.010	&0.992$\pm$0.000	&0.800$\pm$0.003	&0.975$\pm$0.003	&0.934$\pm$0.004	&0.760$\pm$0.005	&0.767$\pm$0.005	&0.876$\pm$0.097	&20.36 \\ 
			LE\_opne	&0.906$\pm$0.011	&0.992$\pm$0.000	&0.803$\pm$0.005	&0.977$\pm$0.001	&0.932$\pm$0.002	&0.764$\pm$0.003	&0.771$\pm$0.006	&0.878$\pm$0.092	&20.00 \\ 
			LLE\_gem	&0.890$\pm$0.008	&0.990$\pm$0.000	&0.704$\pm$0.002	&0.970$\pm$0.004	&0.895$\pm$0.006	&0.726$\pm$0.008	&0.741$\pm$0.005	&0.845$\pm$0.114	&23.57 \\ 
			M-NMF	&0.944$\pm$0.009	&0.992$\pm$0.000	&0.936$\pm$0.001	&0.983$\pm$0.002	&0.983$\pm$0.000	&0.878$\pm$0.008	&0.913$\pm$0.001	&0.947$\pm$0.040	&13.36 \\ 
			AROPE	&0.982$\pm$0.002	&0.991$\pm$0.001	&0.955$\pm$0.001	&0.968$\pm$0.001	&0.967$\pm$0.000	&0.910$\pm$0.001	&0.918$\pm$0.002	&0.956$\pm$0.029	&10.79 \\  \hline
			SDNE\_gem	&\textbf{0.987$\pm$0.004}	&0.979$\pm$0.002	&0.952$\pm$0.000	&0.945$\pm$0.002	&0.971$\pm$0.001	&0.910$\pm$0.002	&0.918$\pm$0.001	&0.952$\pm$0.028	&13.29 \\ 
			SDNE\_opne	&0.985$\pm$0.002	&0.987$\pm$0.000	&0.953$\pm$0.000	&0.957$\pm$0.007	&0.969$\pm$0.002	&0.898$\pm$0.005	&0.917$\pm$0.001	&0.952$\pm$0.032	&13.86 \\ 
			PRUNE	&0.901$\pm$0.010	&0.838$\pm$0.002	&0.956$\pm$0.000	&0.836$\pm$0.003	&0.874$\pm$0.001	&0.904$\pm$0.003	&\textbf{0.920$\pm$0.001}	&0.890$\pm$0.042	&17.71 \\ 
			VERSE	&0.935$\pm$0.010	&\cellcolor{Gray}\textbf{0.994$\pm$0.001}	&\textbf{0.956$\pm$0.002}	&\cellcolor{Gray}\textbf{0.990$\pm$0.002}	&\cellcolor{Gray}\textbf{0.996$\pm$0.000}	&\textbf{0.919$\pm$0.002}	&0.919$\pm$0.002	&\textbf{0.959$\pm$0.033}	&\cellcolor{Gray}\textbf{4.57} \\  \hline
			LINE	&\textbf{0.963$\pm$0.004}	&0.993$\pm$0.001	&0.931$\pm$0.000	&\textbf{0.984$\pm$0.002}	&\textbf{0.991$\pm$0.000}	&0.877$\pm$0.002	&0.882$\pm$0.002	&0.946$\pm$0.048	&12.64 \\ 
			LINE\_opne	&0.850$\pm$0.010	&0.991$\pm$0.000	&0.932$\pm$0.000	&0.933$\pm$0.001	&0.963$\pm$0.001	&0.895$\pm$0.002	&0.894$\pm$0.003	&0.923$\pm$0.045	&19.29 \\ 
			CNE	&0.946$\pm$0.009	&\textbf{0.994$\pm$0.000}	&\cellcolor{Gray}\textbf{0.967$\pm$0.001}	&0.980$\pm$0.000	&0.976$\pm$0.000	&\cellcolor{Gray}\textbf{0.928$\pm$0.001}	&\textbf{0.922$\pm$0.001}	&\textbf{0.959$\pm$0.026}	&\textbf{6.00} \\ 
			\bottomrule
		\end{tabular}
	}
\end{table*}

In this section we present the results of our empirical study. More specifically, in Section~\ref{ssec:lpacc} we discuss LP accuracy, in Section~\ref{ssec:nstruct} the relation between method performance and network structure, in Section~\ref{ssec:ht} hyperparameter tuning, embedding dimensionality in Section~\ref{ssec:embdim}, train-test splits in Section~\ref{ssec:trtef}, edge sampling in Section~\ref{ssec:esplit} and finally binary classification in Section~\ref{ssec:bincl}. All experiments were conducted on a machine equipped with two 12 Core Intel(R) Xeon(R) Gold processors and 256GB of RAM. 

\subsection{Link Prediction Accuracy}\label{ssec:lpacc}

We start in Table \ref{table:LP1} by presenting the AUC-ROC scores for each method on the LP task. These results were obtained using setup LP1 where we additionally tuned the embedding dimensionality, $d \in \{8, 32, 128\}$. All methods performed best for $d=128$, except CNE, which performed best for $d=8$ (see Section~\ref{ssec:embdim} for detailed discussion). The methods are grouped in the table according to the taxonomy in Section~\ref{sec_methods} with the best performing method per group highlighted in bold and the overall best for each network on grey background. The last two columns show the average method performance and average ranking among other methods over all networks. 

The results in Table \ref{table:LP1} show the excellent performance, on most datasets, of the LP heuristics and in particular of RAI and our proposed NE\_heuristics. The latter, together with GraRep, achieves the highest average AUC-ROC amongst all methods of $0.963$. In addition, although the NE\_heuristics method is not a top performer on any particular network, its AUC-ROC scores are consistently high across all networks and never more than 2\% lower than the best scores in each case. This effect, is also reflected by the method's excellent average AUC-ROC rank ($2^{nd}$ overall). Amongst random walk approaches, the implementations of DeepWalk and Node2vec evaluated provide the best results. GraRep is the matrix factorization-based method with highest scores and presents a consistent performance across all networks. Nonetheless, AROPE and HOPE also exhibit good performance. VERSE and CNE are respectively the best neural architecture-based and probabilistic methods. Overall, according to their average AUC-ROCs and AUC-ROC ranks, the best performing methods are: VERSE, NE\_heuristics, GraRep and CNE. It is worth noting that for these top performers we tuned at most one hyperparameter and, thus, their excellent performance can not be ascribed simply to an exhaustive model selection process. Also noteworthy is the fact that CNE achieves state-of-the-art results with an $8$-dimensional embedding as compared to the 128 dimensions required by other methods.

In the absence of additional data, such as node or edge attributes, our experiments suggest an overall improvement in performance as the order of proximity between nodes, captured by a model, increases. GraRep, the best performing NE method, captures up to 8-step relations. HOPE and AROPE, also top performers, capture up to 4-step relations while the remaining first and second order methods, present a lower performance. Exception to this pattern can be found for VERSE and CNE. In the case of VERSE, the second-order method employs a nonlinear transformation able to preserve more information from the original network. CNE, on the other hand, finds an embedding based on first-order information. Its excellent performance can be explained by the fact that it additionally models structural information (node degrees) in the prior, leaving more flexibility to the embedding.

Another important observation from the results in Table~\ref{table:LP1} is that method performance varies significantly between different implementations of the same NE methods (up to 11,7\% for LINE on StundentDB). These differences are especially large for implementations of LINE (3.3\% difference, on average, over all networks), HOPE (2.4\%) and Node2vec (2\%). The main factors causing these differences are numerical instability, approximation of computations and dependencies on different software versions with varying default parameters. In our \href{https://bitbucket.org/ghentdatascience/nrl4lp-public/src/master/}{supplementary material} we also show fluctuations in method performance of up to 44.5\% for \texttt{GEM} implementations due to package dependencies and up to 5\% for metapath2vec due to multi-core execution. Also interesting is the difference in performance between Node2vec and DeepWalk. Here, one would expect the first, which is a generalization of the latter, to perform better. This is indeed the case when the Skip-Gram model is approximated via negative sampling for both methods, as showed by the authors of \cite{grover2016node2vec}. However, in our experiments, the Skip-Gram is approximated via hierarchical softmax for DeepWalk (as proposed by the original authors) and via negative sampling for Node2vec (also as proposed by the original authors). When comparing, for both methods, these two approximations of the Skip-Gram model, we observed a significant difference in LP AUC-ROC of $0.02$ with a standard deviation of $0.02$. 

\subsection{Method Performance and Network Structure}\label{ssec:nstruct}

The results in Table \ref{table:LP1} also show that most NE methods and heuristics perform well on the Facebook, AstroPH and GR-QC networks (a graphical representation of the results in Table~\ref{table:LP1}, is provided in the rightmost heatmap in Fig.~\ref{fig:exp1dim}\footnote{With the exception of CNE presented in Fig.~\ref{fig:exp1dim} for $d=128$ and in Table~\ref{table:LP1} with $d=8$.}). These networks, share similar topologies with large diameters and high clustering coefficients which indicate the presence of community structures within the networks. These observations are consistent with the data represented by the networks i.e. groups of friends in Facebook and citations between papers in different subfields for AstroPH and GR-QC. The community structures with high inter-community edge densities and low intra-community densities could explain the higher LP performance of most methods.

On the other hand, the lowest method performance can be observed for StudentDB, BlogCatalog, PPI and Wikipedia. For StudentDB, the k-partite structure of the network poses a challenge to both heuristics and NE methods. Firstly, due to the fact that in this case similar node neighbourhoods do not necessarily imply that two nodes should be connected. For instance, two students following the same courses will have identical node neighbourhoods, yet, they should never be connected, as links between nodes of the same types do no exist in the data. Also, NE methods must at the same time represent similarity between nodes of different types and dissimilarity for those of the same type. For example, students following the same course must be embedded close to said course yet far from each other.  
Methods able to capture high order proximities between nodes (e.g GraRep, HOPE, AROPE) or those that learn node roles (e.g. Metapath2vec) maintain high AUC-ROC scores in this case. The BlogCatalog and Wikipedia networks present very similar structures with small diameters, high clustering coefficients and large average degrees. This can result in cluttered representations where all nodes are close-by making the LP task harder. Finally, the PPI network presents the opposite scenario with a large diameter and small clustering coefficient. This can result in embeddings where most nodes lie equally far from each other and thus lower LP AUC-ROCs.

\begin{figure*}[ht!]
	\centering
	\begin{minipage}{0.48\textwidth}
		\centering
		\includegraphics[height=3.3cm]{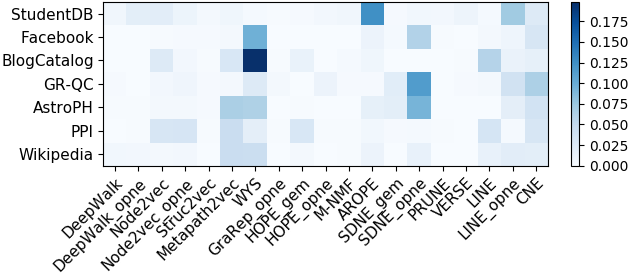}
		\caption{Improvement in AUC-ROC of tuning model hyperparameters. Only methods with tuned parameters are shown.}
		\label{fig:exp0-1}
	\end{minipage} \hskip 0.15in
	\begin{minipage}{0.48\textwidth}
		\centering
		\includegraphics[height=3.3cm]{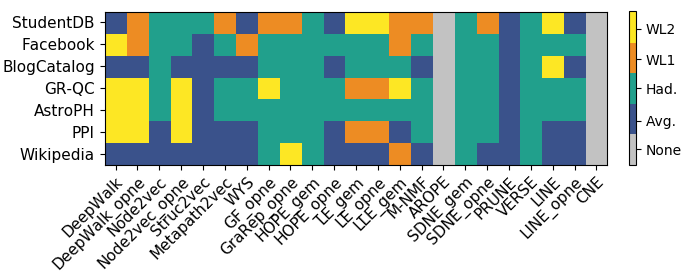}
		\caption{Best performing node-pair embedding operators for each method and datasets on setup LP1.}
		\label{fig:exp1ee}
	\end{minipage}
\end{figure*}

We further analyse the reasons behind the large variations in method performance on different networks by computing the average true positive rates (TPR) and false positive rates (FPR) over all methods for each network. As threshold, we use the value in the AUC-ROC curve that maximises the accuracy of predictions. We observe that while most TPRs at this threshold vary around $0.8$, the FPRs for StudentDB, BlogCatalog, PPI and Wikipedia are up the three orders of magnitude higher that those of other networks. This indicates that, indeed, a clear community structure simplifies the LP task while cluttered graphs negatively impact the prediction of non-link.

\subsection{Hyperparameter Tuning}\label{ssec:ht}

Fig.~\ref{fig:exp0-1} presents a heatmap of the increment in LP accuracy obtained through hyperparameter tuning, i.e. the difference in AUC-ROC between setups LP1 and LP2. Only methods for which hyperparameters were tuned are shown in this figure. The results reveal only limited improvements in most cases, however, in 8\% of the network-method combinations significant improvements of up to 5\% can be observed. Methods including WYS, SDNE and CNE benefit most from hyperparameter tuning. Popular random walk methods such as DeepWalk and Node2vec, for which parameter tuning is tedious, show minimal improvements in AUC-ROC.

In Fig.~\ref{fig:exp1ee} we present the node-pair embedding operator selected using hyperparameter tuning for each method and network on experimental setup LP1. The results indicate that Hadamard is the most frequently selected and that the neural network-based models prefer specific operators, Hadamard in the case of SDNE and VERSE and Average for PRUNE. The remaining methods present a mix of operators. On average, over all methods and datasets we observe a difference in validation AUC-ROC between the best and the worst performing operator of $0.144$ with a standard deviation of $0.091$. This clearly highlights the need to tune the node-pair embedding operator as a method hyperparameter.

\begin{figure}[!ht]
	\centering
	\includegraphics[width=8cm,height=7.7cm]{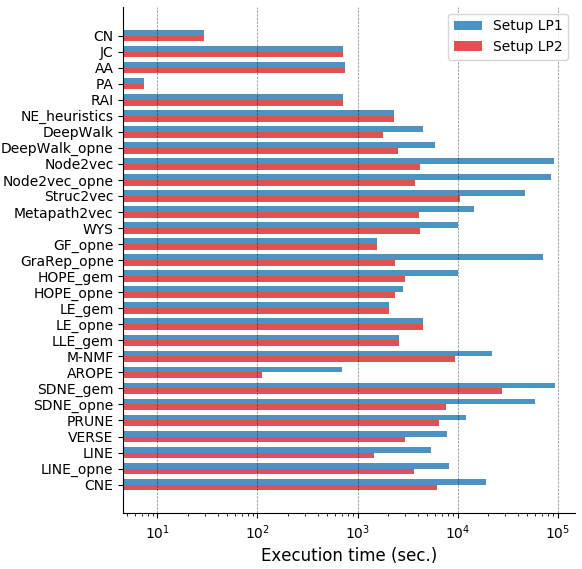}
	\caption{Execution times in seconds of setups LP1 and LP2.}
	\label{fig:exp0-1times}
\end{figure}

For each method, the sum of execution times on the 7 evaluated networks for experimental setups LP1 and LP2 are presented in Fig.~\ref{fig:exp0-1times}. As expected, runtimes for LP1 are larger than those of LP2 and the differences are especially significant for the methods with more tuned hyperparameters e.g. Node2vec, SDNE, AROPE (as this implies computing an embedding of the validation graph for each combination of hyperparameters). Surprisingly, however, the highest increase in runtime can be found for GraRep, for which only the \textit{k-step} parameter was tuned. This indicates that computing high order proximities in GraRep is expensive and, as shown by Fig.~\ref{fig:exp0-1}, does not result in a large improvement in performance. We also observe that naive sequential implementations of the LP heuristics are still faster than heavily optimized and parallelised NE methods with only AROPE presenting comparable runtimes. Finally, with no parameter tuning, similar patterns can be observed within each method family with factorization methods being the fastest.

\subsection{Embedding Dimensionality} \label{ssec:embdim}

\begin{figure}[t]
	\centering
	\includegraphics[width=8cm,height=7.6cm]{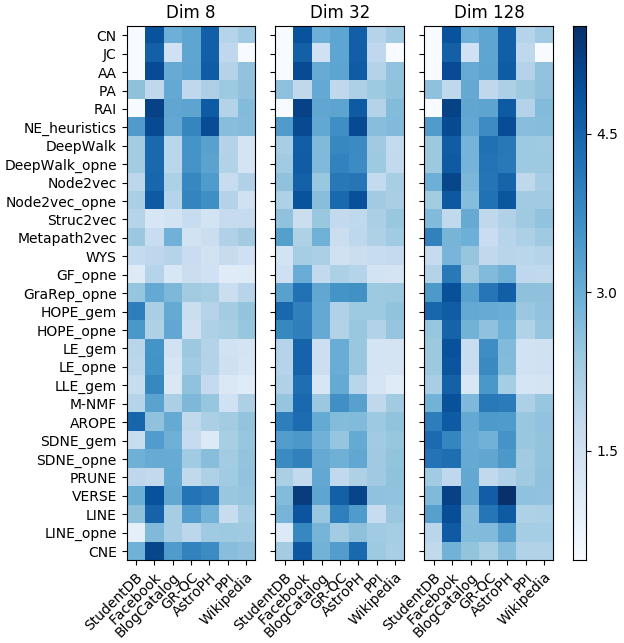}
	\caption{Method performance as $-\log(1-AUCROC)$ on setup LP1 for $d \in \{8, 32, 128\}$. Darker colours indicate better performance.}
	\label{fig:exp1dim}
\end{figure}

We evaluate the effect of embedding dimensionality on method performance by modifying setup LP1 and computing the AUC-ROC of all methods for $d \in \{8, 32, 128\}$. For visualization purposes we compute $-\log(1-AUCROC)$ and present the results in three heatmaps in Fig.~\ref{fig:exp1dim} where a darker colour indicates better performance. The LP heuristics do not depend on $d$ but are shown for reference. These results support the conclusions extracted from Table \ref{table:LP1} regarding the best performing methods. They also show improved performance as $d$ grows for all methods except CNE for which the opposite is true. For AROPE we can observe overfitting on the small StudentDB network. In this case, the method computes embeddings by eigen-decomposition of $\mathbf{A}^3$, which performs well for low values of $d$ but significantly worse as $d$ increases. 

\subsection{Train-Test Split Size}\label{ssec:trtef}

\begin{figure}
	\centering
	\includegraphics[width=8cm,height=7.6cm]{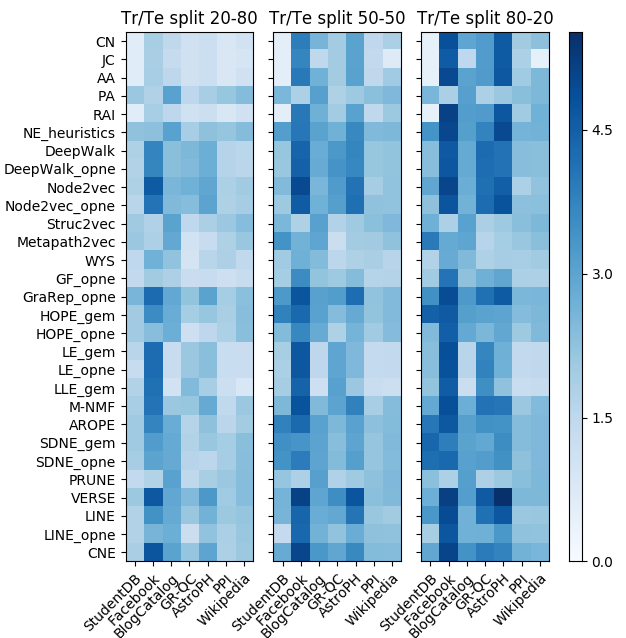}
	\caption{Method performance as $-\log(1-AUCROC)$ on setup LP1 for train-test edge splits 20-80, 50-50 and 80-20. Darker colours indicate better performance.}
	\label{fig:exp2}
\end{figure}

We further use experimental setup LP1 to analyse the effect on method performance of the train fraction $f$. These results are also summarized as three heatmaps in Fig.~\ref{fig:exp2} showing method performance for $f \in \{0.2, 0.5, 0.8\}$. For GR-QC performing a 20-80 split while keeping the training graph connected was not possible due to a low edge density, thus, we resorted to using a 35-65 split instead. An interesting observation from these results is that for $f>0.5$ most methods capture well the network structures while this is not the case for $f<0.5$. This is reflected by an average increase in AUC-ROC over all methods of 7\% between $f=0.2$ and $f=0.5$ while only a 2\% difference can be observed between $f=0.5$ and $f=0.8$. Additionally, we find that neural network-based NE methods are most robust to varying train sizes followed by probabilistic, random walk-based, matrix factorization, and finally the LP heuristics. The proposed NE\_heuristics method, GraRep, VERSE, and CNE best capture the network structure when only 20\% of train edges are available. Lastly, regarding method runtimes, these increase by approximately 50\% from evaluations with $f=0.2$ to $f=0.8$. Some notable exceptions are Metapath2vec, SDNE and Node2vec\_opne, for which the runtimes increase 82.6\%, 78.3\% and 73.2\% respectively. 

\subsection{Edge Sampling} \label{ssec:esplit}

We also conduct an experiment to compare the three strategies introduced in Subsection \ref{sec_lp}, i.e. random, ST and DFT for splitting $\mathbf{E}$ into $\mathbf{E}_{train}$ and $\mathbf{E}_{test}$. Our aim is to study the impact of these strategies on LP performance and sampling execution time. We isolate the effect of this pipeline component by using experimental setup LP2 where no method hyperparameters, apart from the node-pair embedding operator, are tuned. Our results show minimal differences in method accuracy between the three strategies. More precisely, the average AUC-ROC over all methods and datasets using random edge split is $0.897$ with a standard deviation of $0.107$ while for the ST strategy the average AUC-ROC is $0.902$ with the same standard deviation of $0.107$. For DFT the average AUC-ROC is also $0.902$ with a standard deviation of $0.123$.     

\begin{figure}[t]
	\centering
	\includegraphics[width=7.7cm,height=4.3cm]{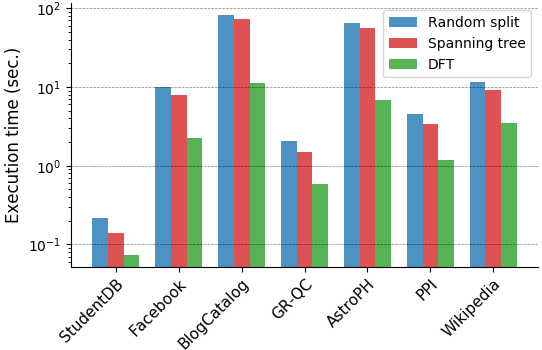}
	\caption{Execution times of different train-test split algorithms.}
	\label{fig:exp5times}
\end{figure}

For very large networks, the edge sampling strategy can become a bottleneck and, thus, sampling runtimes are also worth investigating. Our results, depicted in Fig.~\ref{fig:exp5times}, show that the random strategy is the slowest followed by ST and finally, DFT which is up to one order of magnitude faster on specific networks. The slower runtimes of the random strategy are mainly due to set intersections to obtain the correct $\mathbf{E}_{train}$ and $\mathbf{E}_{test}$. In addition, we have found that the random edge split strategy does not preserve all nodes from the input graph $\mathbf{G}$ in the training graph $\mathbf{G}_{train}$. On average, over all datasets 2.5\% of the nodes in $\mathbf{G}$ are lost. This effect is especially severe for the networks with the lowest average degrees, i.e. StudentDB and GR-QC, which lose up to 8\% of their nodes. The ST and DFT strategies, on the other hand, preserve all nodes.

We have also monitored the variation in performance when embeddings were learned from different initial training graphs $\mathbf{G}_{train}$. For each edge split strategy we have run 3 different executions with varying initial random seeds and in all cases we used an 80-20 train-test split. We did not find any significant differences with a maximum standard deviation observed across all datasets, methods and strategies of only $0.004$. The largest variations were observed for the smallest network, StudentDB. Ultimately, this experiment reveals that averaging LP results over several sets $\mathbf{E}_{train}$ and $\mathbf{E}_{test}$ generated using different random seeds ---in order to obtain unbiased estimates of method performance--- becomes less necessary as the train network sizes ($\mathbf{G}_{train}$) surpass a few thousand nodes. 

\begin{figure}
	\centering
	\includegraphics[width=8cm,height=4.7cm]{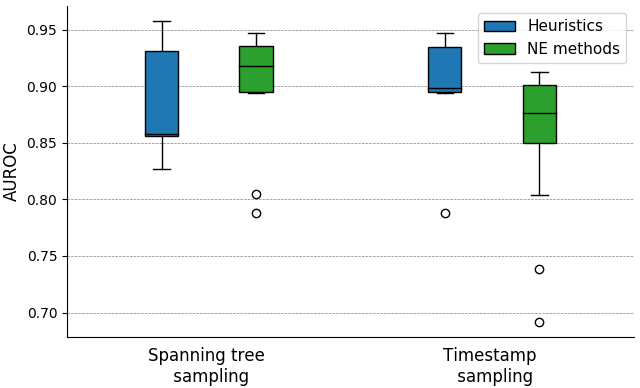}
	\caption{Performance of the heuristics and NE methods compared for the spanning tree and timestamp-based sampling strategies.}
	\label{fig:exp5tsVSst}
\end{figure}

Finally, we compare the ST strategy for randomly splitting edges in sets $\mathbf{E}_{train}$ and $\mathbf{E}_{test}$ to a split based on edge timestamps. This setup will also allow us to determine if the excellent performance of the heuristics is boosted by the random edge sampling strategy used. For this, we use two temporal networks \href{https://snap.stanford.edu/data/CollegeMsg.html}{CollegeMsg} and \href{https://snap.stanford.edu/data/sx-mathoverflow.html}{MathOverflow} obtained from the SNAP repository \cite{leskovec2015snap}. The first network encodes messages sent between students at UC Irvine while the second captures interactions between users of the Math Overflow social platform. As both networks present temporal information on the edges, this allows us to compare a timestamp-based edge sampling to the random ST strategy. We do this by using setup LP2 with a fixed train fraction of $0.8$. In the timestamp edge sampling we select the 20\% most recent edges such that, when these are removed, the remaining graph is still connected. Our results are summarized in Fig.~\ref{fig:exp5tsVSst}. Firstly, we observe opposite effects on method performance for the two sampling strategies. While the random ST sampling appears to boost the performance of NE methods over that of the heuristics, the timestamp-based sampling favours the performance of the heuristics. If we now compare heuristic performance between the two strategies, we observe somewhat similar AUC-ROC scores. For the NE methods, however, performance clearly improves with the random strategy. This indicates that the random sampling strategies are, in fact, boosting the performance of the NE methods over that of the heuristics and not vice versa.

\subsection{Binary Classification} \label{ssec:bincl}

\begin{figure}
	\centering
	\includegraphics[width=9cm]{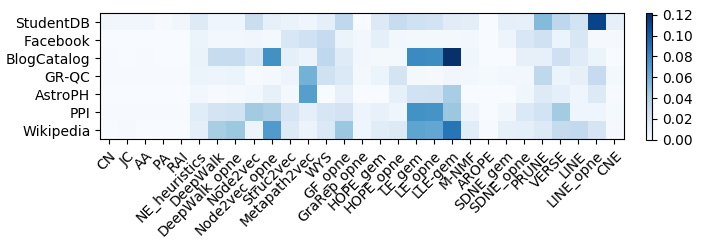}
	\caption{Standard deviations in LP AUC-ROC with different binary classifiers.}
	\label{fig:exp6}
\end{figure}

In this experiment we evaluate the changes in AUC-ROC due to the binary classification method used. We adapt experimental setup LP2 and consider the following classifiers: logistic regression (LR), logistic regression with 5-fold cross validation (LRCV) and decision trees (DT). A heatmap summarizing the standard deviations of the AUC-ROC scores, obtained for each method and dataset, using the three aforementioned classifiers, is presented in Fig.~\ref{fig:exp6}. In contrast to recent results suggesting large variations in performance between LR and LRCV \cite{Gurukar2019}, we did not find significant differences in our evaluation. The large standard deviations in the figure are fundamentally due to the DT classifier. When predictions were given by a DT classifier, the performance of two matrix factorization approaches, LE and LLE, improved. For LE we observed an increment in average AUC-ROC over all networks of $6.5$\%, from $0.87$ using LRCV to $0.93$ with DT. For LLE the increment was of $7.8$\%, from an AUC-ROC of $0.84$ with LRCV to $0.91$ with DT. All other methods, and particularly the \texttt{OpenNE} implementations of Node2vec and LINE, showed lower AUC-ROC scores for DT as compared to LRCV. 

\subsection{External validation of the results}
Due to differences in the evaluations, we can only validate our results, to some extent, against the empirical results in the Node2vec and CNE papers. In the first case, although our AUC-ROC scores are consistently higher, the same main conclusion holds: node2vec outperforms its competitors on the three networks originally selected. However, a properly tuned LINE exhibits similar performance or even outperforms node2vec on other networks. In the second case, our results are very similar to those reported by the authors of CNE. Finally, our evaluation also corroborates the conclusions reported by the authors of AROPE and VERSE as well as the choice of node-pair operator for the latter.

\section{Conclusions and Discussion}\label{sec_conclusions}

In this paper we have conducted an extensive empirical study on NE methods for LP. Our results show that, despite the surge of interest in the field in recent years, thin progress has been made. Complex state-of-the-art NE methods can be matched or even outperformed by simple baselines with no hyperparameters, in terms of both accuracy and runtime. On average, the proposed NE\_heuristics baseline outperforms all others methods for the LP task. Nevertheless, some embedding-based methods such as GraRep, VERSE and CNE do show promising results.
We have also shown that clear community structures in the graphs, high embedding dimensionalities and capturing high order proximities between nodes all impact method performance positively. In contrast with recently published results, our experiments indicate no significant difference between LR and LRCV as binary classifiers in this setting. We also highlight the need to tune the node-pair embedding operator as model hyperparameter and that a single train-test split provides a good estimation of method accuracy resulting in higher evaluation efficiency. Finally, we hope this study and evaluation toolboxes such as \texttt{EvalNE} serve as initial steps towards the creation of standard evaluation pipelines for NE methods, and shed light on the current state-of-the-art. Specifically, we plan to use these results as a basis for an online resource that is continuously augmented with results for newly developed methods and/or other networks.

\bibliography{bibliography}
\bibliographystyle{ieeetr}

\end{document}